\documentclass[twocolumn,nofootinbib,preprintnumbers,amsmath,amssymb]{revtex4}

\usepackage{graphicx}
\usepackage{dcolumn}
\usepackage{bm}
\usepackage{amsmath}
\usepackage{color}

\def \be {\begin{equation}}
\def \ee {\end{equation}}
\def\bea{\begin{eqnarray}}
\def\eea{\end{eqnarray}}
\begin{document}

\title{Gravitational Radiation from Massless Particle Collisions}

\author{Andrei Gruzinov${ }^{1}$ and Gabriele Veneziano${ }^{1,2,3}$}

\affiliation{{\sl ${ }^{1}$ CCPP, Physics Department, New York University, 4 Washington Place, New York, NY 10003} \\
{\sl ${ }^{2}$Coll\`ege de France, 11 place Marcelin  Berthelot 
 75005 Paris, France}\\ 
{\sl ${ }^{3}$Theory Division, CERN, CH-1211 Geneva 23, Switzerland
}}

\begin{abstract}

  We compute  classical gravitational bremsstrahlung from the gravitational scattering of two massless particles at leading order in the (center of mass)  deflection angle $\theta\sim  4 G \sqrt{s}/b = 8 G E/b \ll 1$. The calculation,  although non-perturbative in the gravitational constant, is surprisingly simple and yields explicit formulae --in terms of multidimensional integrals--  for the frequency and angular distribution of the radiation.  In  the range $ b^{-1} < \omega <  (GE)^{-1}$,  the GW spectrum behaves like $ \log (1/GE\omega) d \omega$, is confined to cones  of angular sizes (around the deflected particle trajectories) ranging from  $O(\theta)$ to $O(1/\omega b)$, and exactly reproduces, at its lower end,  a well-known zero-frequency limit.
    At $\omega > (GE)^{-1}$ the radiation is confined to cones of angular size of order $\theta (GE\omega)^{-1/2}$ resulting in a scale-invariant  ($d\omega/\omega$) spectrum. The total efficiency in GW production is dominated by this "high frequency" region and is  formally logarithmically divergent in the UV. If the spectrum is cutoff at the limit of validity of our approximations (where a conjectured bound on GW power is also saturated), the fraction of incoming energy radiated away turns out to be  $\frac{1}{2 \pi} \theta ^2 \log \theta^{-2}$ at leading logarithmic accuracy.
 \end{abstract}
\maketitle

\section{Introduction}

Investigations about gravitational radiation in ultrarelativistic two-body collisions in  Classical General Relativity (CGR) go back to classic work by  D'Eath \cite{death} and by Kovacs and Thorne \cite{KT} in the late seventies (see also \cite{Damour}). In  these papers one considers a pure gravitational collision of two pointlike particles of masses $m_1, m_2$, center of mass energy $E = E_1 + E_2 = m_1 \gamma_1 + m_2 \gamma_2 $ and  impact parameter $b$. The goal is to calculate the frequency  and angular distribution of  the outgoing gravitational radiation.
More modestly, one would like to compute (or at least estimate) the fraction $\epsilon_{GW}$ of the initial energy that gets radiated in GW as a function of the dimensionless parameters of the problem. In the geometric units that we shall use throughout (for which $G=c=1$) these are:  $E_i/b$, (from which one can compute the classical deflection angle $\theta$) and the Lorentz factors $\gamma_i$.  Since this work concerns the massless limit we shall restrict ourselves from the start to the equal mass case.

Unfortunately, the techniques employed in   \cite{death} and \cite{KT} can only be fully trusted at $\theta < \gamma^{-1}$ i.e. in a range of scattering angles that shrinks to zero as the particles become ultra-relativistic. Although in \cite{death} formal expressions are given for generic $\gamma$, not much is done in the way of taking the large-$\gamma$ limit at fixed (even if small) $\theta$. Nothing could be said, in particular, about the  gravitational scattering of strictly massless particles.
This unfortunate limitation was recently confirmed in a series of papers by Gal'tsov et al. \cite{Gal'tsov}. These authors found agreement  with  \cite{death},  \cite{KT} in the same (very small angle) regimes  -and managed to extend the results to the case of extra spatial dimensions- but still only sufficiently away from the massless limit.
The large $\gamma$ limit is also numerically challenging \cite{numerical} since it is difficult to disentangle the emitted GW from the shock-wave metric of the two ultra-relativistic particles after their encounter since they both travel with the speed of light. This is to be contrasted with the so-called hyperbolic (i.e. no-merger) non-relativistic black-hole encounters for which both numerical relativity and effective-one-body  techniques can be successfully applied \cite{EOB}.  In so far as those results can be compared with ours no obvious disagreement is found.

Although not practically pressing, the problem has an obvious methodological value, and can be used as a test for numerical general relativity. Besides, it appears naturally in studies of the trans-Planckian-energy quantum collision of massless strings among themselves  \cite{ACV} and with branes \cite{DDRV}. In this regime gravity dominates and various CGR results, such as gravitational deflection and tidal excitation, have been reproduced.  Concerning gravitational bremsstrahlung the expected flat spectrum at low graviton-energies was also recovered \cite{ACV, VW}. However, technical difficulties were encountered in estimating the (order of magnitude of the) cutoff on the frequency/energy of the emitted gravitons. The most naive estimates gave $\omega < \omega_{cr} \sim \theta ^{-2}  E^{-1}$  leading to an overproduction of relatively energetic gravitons if the low-energy flat spectrum is extrapolated up to $\omega_{cr}$. This would push $\epsilon_{GW}$ to become $O(1)$ already for $\theta \sim M_P/E \ll 1$. Precisely in this regime, many gravitons are produced and a classical limit, if it exists, should be recovered. This observation partly motivated the present study.

Of course, one can also imagine the massless (or $\gamma \rightarrow \infty$) limit to be singular. However, since the classic paper by Weinberg \cite{Weinberg}, we know that gravitational bremsstrahlung, unlike its electromagnetic counterpart, does not suffer from collinear singularities, i.e. has a smooth massless limit. 
In this short note we shall argue that, indeed, the massless limit exists and,  under some reasonable assumption, is easily calculable resulting in a simple, elegant result. 
Actually, we have shown that equations (6.17)-(6.19) of \cite{death} do have a smooth $\gamma \rightarrow \infty$ limit:  our calculation is way easier than that of \cite{death}; in fact, we think that our approach justifies, a posteriori,  taking the formal  $\gamma \rightarrow \infty$ limit of  D'Eath's result. 

Though simple, our derivation is somewhat subtle and involves an effective resummation in powers of Newton's constant, perhaps explaining why previous methods have failed. It has nonetheless the shortcoming of neglecting non-linear effects in a strong-curvature regime forcing us to exclude a small-distance, high-frequency region from consideration. In principle, a more rigorous and powerful method for settling the question of this region's contribution can be based on the so-called radiation moment approach developed by Vilkovisky 
\cite{Grisha}. Unfortunately, we have not be able, so far, to take full advantage of it.

In Section II we present our derivation of an expression for the differential GW spectrum in terms of a two-dimensional integral. In Section III we present an analytic treatment of the frequency spectrum after  approximately integrating over emission angles, while in Section IV we present a more detailed and rigorous -but also more qualitative- study of the angular and frequency distribution.  Section V contains our conclusions and outlook.

\section{The calculation}

We will calculate gravitational bremsstrahlung in the CM frame, where two particles of energy $E$ collide at an impact parameter $b\gg E$. 
Before the collision, by causality, the metric is given by the superposition of two non-interacting  Aichelburg-Sexl (AS) shock waves \cite{aich} while, because of their compact support, each wave is deflected and distorted by the other one in a calculable way (in a suitable approximation) after the collision\footnote{A similar approach has been used \cite{Headon} to study the head-on collision of two massless particles leading to black-hole formation with a finite fraction of the incoming energy emitted in GWs.}.

We shall need the following inputs: (i) the AS metric of a massless particle and its Riemann tensor, (ii) the null geodesics in this metric, (iii) the energy flux of the planar gravitational wave in terms of the Riemann tensor of the wave,  (iv)  Huygens principle in the Fraunhofer limit. We address these issues in turn.

(i) The AS metric of a massless particle of energy $E$ is:
\be\label{metric}
ds^2=-dz^+dz^-+dx^2+dy^2-4E\ln \frac{x^2+y^2}{\lambda^2} \delta (z^+)(dz^+)^2.
\ee
Here $z^{\pm}=t\pm z$, and the particle's world line is $x=y=z^+=0$\footnote{For dimensional reasons, the metric  contains an arbitrary length scale $\lambda$. By consistency, and as a check of our results,  $\lambda$ will disappear from all observable quantities.}.  
 
The relevant components of the Riemann tensor of (\ref{metric}) are given by
\be\label{riemann}
R \equiv  R_++iR_\times =4E\delta(z^+){\zeta^2\over |\zeta|^4}
\ee
(not to be confused with the Ricci scalar), where 
\be
R_+\equiv {1\over 2}(R_{+x+x}-R_{+y+y}),~~R_\times \equiv R_{+x+y} \, ,
\ee
the $+$ indices on the r.h. sides stand for $z^{+}$,
and we have introduced  a convenient complex notation with $\zeta = x + i y$.
Note that $R$ is independent of $z^-$ and is also an (anti)analytic function on $\zeta$. As a result it obeys the free wave equation:
\be
\label{waveeq}
\Box R \equiv 4 (- \partial_+ \partial_- + \partial_{\zeta} \partial_{\bar{\zeta}}) R = 0\, .
\ee

(ii) Null geodesics of normal  incidence (with respect to the gravitational shock eq.(\ref{metric})) are given by the following (exact) expressions (see e.g. \cite{FPV}): 
\bea
x&=&a-{4E\over a}z^+ \theta (z^+), ~~y=0, \nonumber \\
~~z^-&=& - \left(8E\ln (a/\lambda) -  \left({4E\over a}\right)^2z^+ \right) \theta (z^+),
\eea
where we assumed the geodesic to be at $z^-=0$ before hitting the shock wave at an  impact parameter $a$. This world line corresponds to a deflection angle
\be 
\label{angle}
\theta (a) = 2 \tan^{-1}(4E/a),
\ee
and to $t$ and $z$ delays given by $\delta z^+ =0$ and by
\be\label{phase}
\delta z^-= - 8E\ln (a/\lambda) \Rightarrow \delta t = - \delta z =  - 4E \ln (a/\lambda).
\ee

(iii) The energy flux density of gravitational waves at future null infinity ${\cal I}^+$  is completely determined by the Bondi-Sachs complex news function $\partial_u C$:
\be
 \frac{d E^{GW}}{du} = \frac{1}{2\pi} \int d^2 \Omega |\partial_u C|^2 \, ,
\ee
where $\Omega$  and $u$ are, respectively,  the solid angle and  retarded time (an exactly null coordinate in the Bondi-Sachs parametrization of the metric). At ${\cal I}^+$, $u   \rightarrow (t-r)$. In turn, the  news function at ${\cal I}^+$  is related to the asymptotic form of the Riemann tensor by:
\be
\frac{\partial^2}{\partial u^2} C = - r R ~~;~~ r \rightarrow \infty \, .
\ee
It is convenient (and also  physically more interesting) to go from $u$ to ``frequency"  $\omega$ by a Fourier transform. One then easily gets:
\be
\label{flux}
		\frac{d E^{GW}}{d\omega} ={1\over   \omega^2}  \int d^2 \Omega~ r^2  |{\cal R}_{{\cal I}^+} |^2 \, ,
\ee
where  ${\cal R}_{{\cal I}^+}$ is the Fourier transform of  $R$ at ${\cal I}^+$.

(iv) Finally,  to the future of the collision plane, $R$ satisfies an equation of the form (see e.g. \cite{GAVnotes}):
\be
\label{waveeq1}
\Box R + O(R^2) = 0\, ,
\ee
i.e. it obeys again (\ref{waveeq}) modulo higher order corrections which are quadratic in the curvature. As long as the latter can be neglected (see below for a discussion on this restriction)
we can apply Huygens principle in the Fraunhofer  approximation (all characteristic distances must be much greater than the wavelength and angles w.r.t. the direction of the incident wave must be small, see e.g. \cite{LL}) to write a solution  of the above equation. Concentrating on the forward hemisphere, it will give ${\cal R}({\bf R})$  at a far-away point ${\bf R}= r (\hat{z}+{\boldsymbol \rho })$ (with ${\boldsymbol \rho } \perp \hat{z}$, $\rho \ll 1$ representing the  small angle at which the GW arrives on ${\cal I}^+$) in terms of the wave amplitude $ {\cal R} ({\bf x})$ at the ``screen" situated at $z=0$, with ${\bf x} \perp \hat{z}$:
\be\label{huygens}
r {\cal R} ({\bf R})={\omega \over 2\pi  i }\int d^2 {\bf x}~{\cal R}({\bf x})e^{-i\omega u({\bf x}, {\boldsymbol \rho })}.
\ee

Here $u({\bf x}, {\boldsymbol \rho })$ is the asymptotic value of $u$ for a null geodesic starting from ${\bf x}$ and going to  ${\cal I}^+$ in the direction ${\boldsymbol \rho }$. The null character of $u$ insures that (\ref{huygens}) is indeed a solution of (\ref{waveeq}).  Since an overall (i.e. $x$-independent) phase is irrelevant, what really matters is the difference between the values of $u$ (in a given direction $\rho$) for null geodesics coming from different points $x$ on the screen. Finally, since to a very good approximation $u = z^-$ in these nearly forward directions we can take  ${\cal R}({\bf x})$ to be the Fourier transform with respect to $z^-$ of  the curvature $R$ for the right-moving AS wave (analog of (\ref{riemann}) but with $z^+$ replaced by $z^{-}$) at $z = t = 0$ i.e.:
\be
\label{FTR}
{\cal R}({\bf x}) = \frac{4E}{2 \pi} {\zeta^2\over |\zeta|^4}\, .
\ee

With all these ingredients we now  calculate the energy emitted close to the positive-$z$ direction in the collision of a particle, moving in that same direction and originally at ${\bf r=0}$, after being scattered by a particle moving in the opposite direction and originally at ${\bf r}={\bf b}$. We integrate the flux (\ref{flux}), with the wave (Riemann) amplitude given by (\ref{huygens}), the amplitude at the screen given by (\ref{FTR}), and the phase at the screen given by (\ref{phase}). The result is:
\bea
\label{finalDelta}
E^{GW} &=& {E^2\over  \pi^4}\int d^2\rho~ d\omega |c |^2,  \\
c(\omega, {\boldsymbol \rho }) &=&\int d^2 x ~e^{-i\omega {\boldsymbol \rho }\cdot {\bf x}} \cdot \frac{\zeta^2}{|\zeta|^4} \left( e^{i \omega \Delta z^-} - e^{i \omega  \Delta z^-_{AS}}  \right)  \nonumber \\
 \frac{ \Delta z^-}{4E}&=& - \ln \frac{({\bf x}-{\bf b})^2}{\lambda^2} ;\frac{  \Delta z^-_{AS}}{8E} =  -  \ln {b \over \lambda} + {{\bf b}\cdot {\bf x}\over b^2}\, . \nonumber
\eea

 Eq. (\ref{finalDelta}) requires some explanation. The first exponential in (\ref{finalDelta}) is the  (kinematical) phase factor of the standard (non gravitational) Fraunhofer approximation. Instead,  $\Delta z^-$ originates from the time delay and is peculiar to gravity.
  $\Delta z^-$  is simply the $u$-delay for a generic point on the screen according to (\ref{phase}).
Thus the first exponential in  (\ref{finalDelta}) gives the full  news function at ${\cal I}^+$. However,  part of it has to be assigned to the deflected AS shock wave.
Therefore, following \cite{death}, we have subtracted (see second exponential in  (\ref{finalDelta})) the contribution of an AS wave 
 that travels at an angle  $\theta = {8E\over b}$ (eq. (\ref{angle}) for $a =b$ and  small $\theta$), and which has the retarded time delay given again by (\ref{phase}) plus an additional geometric delay due to the tilt of the deflected front wave. 
 In other words the truly emitted gravitational wave is the difference between the actual Riemann tensor and the Riemann tensor of the deflected/delayed shock wave.
 
 This subtraction removes any dependence of $E^{GW}$ from $\lambda$.
 Indeed, an overall phase $8 E \omega  \ln {b \over \lambda}$ can be pulled out of  (\ref{finalDelta}) to give the equivalent expression: 
\bea
\label{finalc}
&&c(\omega, {\boldsymbol \rho }) =  \\
 && \!\!\!  \int \frac{d^2 x~ \zeta^2}{|\zeta|^4} ~e^{-i\omega {\bf x} \cdot ( {\boldsymbol \rho } -  8E  \frac{{\bf b}}{ b^2} )}
\left[ e^{-4i E \omega ( \ln \frac{({\bf x}-{\bf b})^2}{b^2} + 2  \frac{{\bf b}\cdot {\bf x}}{ b^2})} -
 1 \right] \, . \nonumber 
\eea

Note that, according to (\ref{finalc}),  the $\rho$-dependence of the spectrum (i.e. its angular distribution) depends just on the combination $( {\boldsymbol \rho } - 8E  \frac{{\bf b}}{ b^2})$  which is nothing but the polar coordinates around a deflected ``north pole" defined by the direction of the outgoing particle, something physically sensible.
Inserting (\ref{finalc}) into (\ref{finalDelta}) gives our final result for the angular and frequency distribution of the emitted radiation:

\bea
\label{differential}
\frac{dE^{GW} }{ d\omega~d^2\rho_s} &=& {E^2\over  \pi^4} |c |^2~;~  {\boldsymbol \rho }_s =   {\boldsymbol \rho }- 8E  \frac{ {\boldsymbol b }}{ b^2} \, ;\nonumber  \\
c(\omega,  {\boldsymbol \rho }_s) &=& \int \frac{d^2 x~ \zeta^2}{|\zeta|^4} ~e^{-i\omega {\bf x} \cdot  {\boldsymbol \rho }_s }
\left[ e^{-i E \omega \Phi({\bf x})} - 1 \right] \, ; \nonumber  \\
  \Phi({\bf x})  &=& 4  \ln \frac{({\bf x}-{\bf b})^2}{b^2} + 8  \frac{{\bf b}\cdot {\bf x}}{ b^2}\, , 
\eea
whose  consequences  we will now discuss.

\section{An approximate analytic treatment}

A detailed analysis of the frequency and angular spectrum implied by (\ref{differential}) is quite non-trivial. It will be sketched in the next Section, but its full completion may require a non-trivial numerical analysis. Here we give a simplified treatment aiming at the
frequency spectrum after angular integration. Let us do that after rewriting (\ref{differential}) in the equivalent form:
\bea
\label{EGW}
&&E^{GW} =  {E^2\over  \pi^4}\int d^2\rho_s~ d\omega \int \frac{d^2 x_1 \zeta_1^2}{|\zeta_1|^4}  \int \frac{d^2 x_2 \zeta_2^{*2}}{|\zeta_2|^4}   \\
&& e^{- i \omega{\bf \rho}_s({\bf x_1} - {\bf x_2})}\left( e^{-iE \omega \Phi(x_1)}-1\ \right) \left( e^{+ iE \omega \Phi(x_2)}-1\ \right) \nonumber
\eea
In general integration over a finite range of ${\bf \rho}_s$ would produce some Bessel functions of $|{\bf x_1} - {\bf x_2}|$. However, if the characteristic scale of $\omega$ is $E^{-1}$ and that of $x$ is $b$, the effective integration region should be sufficiently large to allow  replacing the exact result by a $\delta$-function using:
\be
\label{delta}
\int d^2 {\bf \rho}_s~e^{- i \omega{\bf \rho}_s({\bf x_1} - {\bf x_2})} \sim {(2\pi )^2\over \omega^2} \delta^{(2)} (x_1 -  x_2)\, .
\ee
This approximation leads to a very simple analytic result for the frequency spectrum:
\be
\label{spectrum}
 {\omega ^2 dE^{GW}\over d\omega}  ={16 E^2\over \pi^2}\int {d^2 x \over |x|^4} \sin ^2 \left({\omega E \Phi(x)\over 2}\right).
\ee
In the next Section we shall see to what extend this approximation is  justified. For the moment let us proceed as if it was and work out its consequences.

Note that, at any fixed $\omega$, the  integral in (\ref{spectrum}) is convergent both in the IR (large $x$) and in the UV (small $x$). The first property is obvious, while the second follows from the small-$x$ behavior of $\Phi(x)$,  $\Phi \sim -4(Re~ \zeta^2)/b^2$.
However, if we now perform the $\omega$-integration, we get:
\be
\label{omegaint}
E^{GW} ={16 E^3 \over \pi} \int {d^2 x \over |x|^4}|   \ln \frac{({\bf x}-{\bf b})^2}{b^2} + 2  \frac{{\bf b}\cdot {\bf x}}{ b^2}| 
\ee
and the resulting integral diverges logarithmically at small $x$.
With leading-log accuracy we find:
\be
\label{ll}
E^{GW} = - {64 \over \pi}{E^3\over b^2} \ln  |{\bf {x}_{{\rm min}}\over b}| = - { \theta^2 \over \pi} E  \ln  |{\bf {x}_{{\rm min}}\over b}| \,
\ee
implying a UV sensitivity for the total emitted energy. To estimate $\bf {x}_{{\rm min}}$ 
note that, for $|x| < E$,  $R$ becomes large and the $O(\tilde{ R}^2)$ corrections in (\ref{waveeq1}) are  no longer negligible. This strongly suggests using $ |{\bf x}_{{\rm min}}| =  E$, an estimate we shall be using hereafter. Of course, the total GW yield will depend on what actually happens in the strong-curvature region $|x| < E$. Our educated guess, for which a  justification will be given shortly, is that the non-linear terms in $R$ will effectively cutoff the spectrum at high frequency. In order to address better this point,
 let us discuss the relevant region in $|x|$ for different regimes of $\omega$ according to  (\ref{spectrum}):
\begin{itemize}
\item $b^{-1} < \omega < E^{-1}$.
Here the integral in (\ref{spectrum}) is effectively cutoff at $|x| \sim b (\omega E)^{-1} \gg b$ and we get\footnote{This was confirmed by an 
accurate numerical calculation of the integral in(\ref{spectrum}).}:
\be 
\frac{1}{E} {d E^{GW} \over d\omega} = {4 \over \pi} \theta ^2 E
\ln (\frac{1}{ E\omega}) \; .
\label{lowom}
\ee
\item
 $E^{-1} < \omega <  \theta^{-2}  E^{-1}$.
The integral in (\ref{spectrum}) is now cutoff at $|x| \sim b (\omega E)^{-1/2} \ll b$ (but  $\gg E$) giving:
\be 
\label{intermediate}
\frac{1}{E} {d E^{GW} \over d\omega} = {1 \over 2  \pi} \theta ^2  \omega ^{-1} \, ,
\ee
in agreement with (\ref{ll}) for $(\frac{\bf{x}}{b})_{{\rm min}}^2 \sim (\omega E)_{{\rm max}}^{-1}$.
Note that at the upper limit of this range $\frac{1}{E} {d E^{GW} \over d\omega} \sim \theta^4$ i.e. of the same order as the higher-order corrections that we neglected\footnote{By Fourier transforming (\ref{intermediate}) back to $u$ one finds that, at $u \le E \theta^2$, $dE^{GW}/du \ge 1$ violating a widely conjectured bound on GW power (see \cite{GWPB} and references therein). This  problem is neatly avoided if the spectrum cuts off above $\omega =  \theta^{-2}  E^{-1}$ in the way discussed below.}.
\item $\omega >  \theta^{-2}  E^{-1 } $. The spectrum would continue its $\omega^{-1}$ behavior unless
we cutoff the integral  at $|x| \sim E$ (the border of the non-perturbative region, see previous discussion) in which case we obtain:
\be 
 {d E^{GW} \over d\omega} = {16 \over \pi}  \frac{E^2}{\bf {x}_{{\rm min}}^2} \omega ^{-2} \sim {\rm const.}~ \omega ^{-2}   \, .
\ee
\end{itemize}
For $\bf {x}_{{\rm min}} \sim E$ (with a suitable proportionality constant) such a spectrum would coincide  with the time-integrated spectrum of an evaporating black hole, which, amusingly, turns out to be $M$ and $\hbar$-independent at high-frequency ($\omega M \gg 1$).

Concentrating ourselves on the first regime and going to its lower limit  $\omega \sim b^{-1}$ we can connect our result to the so-called ``zero-frequency-limit" 
which can be computed \cite{ZFL} from Weinberg's soft graviton limit \cite{Weinberg} (see also \cite{WeinbBook}):
\be
 {d E^{GW} \over d\omega} \rightarrow - {4 \over \pi} G t 
\ln \left(- \frac{s}{t}\right)~;~ s = 4 E^2 ~,~ \theta^2 \sim -  \frac {4t}{s}\, ,
\ee
or, equivalently:
\be
\label{ZFL}
{d E^{GW} \over d\omega} \rightarrow {4 \over \pi} \theta ^2 E^2
\ln ( \theta ^{-2}) \; ,
\ee
which coincides with the limit $\omega \rightarrow b^{-1}$ of the spectrum (\ref{lowom}) after accounting for the emission from both colliding particles.

\section{A more detailed analysis}

In this section we give an evaluation of $c$ (and consequently of the full differential spectrum) as given in  (\ref{differential}) in terms of a two-dimensional integral.
To this purpose let us  define first some reduced dimensionless quantities:
\be
\tilde{\omega} = E \omega~;~\frac{\zeta}{b} = w e^{i \phi}~;~ {\boldsymbol \delta_s} = \frac{{\boldsymbol \rho_s}}{\theta}~;~\cos(\phi - \psi) = {\bf \hat{x}} \cdot {\bf \hat{\rho}_s} \, ,
\ee
with $\phi$ the angle between ${\bf x}$ and ${\bf b}$ and $\psi$  the angle between ${\boldsymbol \rho_s }$ and ${\bf b}$. Eq. (\ref{differential}) can then be rewritten as:
\be
\label{c1}
c = \int_{w_{min}} \frac{dw}{w} \int_0^{2\pi} d\phi~ e^{2i \phi} e^{-8 i w \tilde{\omega} {\boldsymbol \delta }_s \cdot \hat{{\bf x}}} \left[ e^{-i \tilde{\omega} \Phi(w, \phi)} - 1 \right]\, ,
\ee
where ${\boldsymbol \delta }_s \cdot \hat{{\bf x}} = \delta_s  \cos(\phi - \psi)~;~ \delta_s \equiv  |{\boldsymbol \delta_s}|$ and:
\be
 \Phi(w, \phi) = 4 \left[ \log( 1 - 2 w \cos \phi + w^2) + 2 w  \cos \phi  \right] .   
\ee
Note that the Fraunhofer approximation requires $\omega > 1/b$ ($\tilde{\omega} > \theta$) and $|{\boldsymbol \rho }| \ll 1$ ($\delta \ll \theta^{-1}$).
As already discussed we have justified our neglect of higher order terms in $R$ by also requiring $w >  w_{min}$ with $w_{min} \sim \theta$. 
In terms of the above rescaled variables, the yield in GW can be written as:
\be
\label{EGW1}
E^{GW} = \frac{E \theta^2}{ \pi^4} \int d \tilde{\omega} ~d^2  \delta_s~ |c(\tilde{\omega}, \delta_s, \psi,w_{min})  |^2  \, ,
\ee
which is $O(E~ \theta^2)$ unless the above-mentioned cutoffs in $\omega$, $|{\boldsymbol \rho }|$ and $w$ become essential.

We shall
analyze the behavior of $c$ in two complementary frequency regimes and, for each one of them, we will determine the regions in $w$ and $\delta_s$ that are most relevant for the total GW yield.
\vspace{2mm}

\centerline{  {\bf A: $\tilde{\omega} > 1$ i.e. $\omega > 1/E$}}

We separate the $w$ integral in (\ref{c1}) in two regions: $w_{min} < w < 1$, $w >1$, and claim that the first region gives the leading contribution. In that region we can use $\Phi \sim -4 w^2 \cos 2 \phi$, and, changing variable from $w$ to $\lambda = w  \tilde{\omega}^{1/2}$,  we write:
\be
\label{dEGW}
\frac{dE^{GW}}{d \log \tilde{\omega}~ d \psi}  = \frac{\theta^2 E }{ \pi^4}\int _0^{\delta_s^{max} \tilde{\omega}^{1/2}} \hat{\delta}_s d  \hat{\delta}_s \sum_i |c_i^<|^2
\ee
where  $i = +, \times$,  $ \hat{\delta}_s \equiv \delta_s \tilde{\omega}^{1/2}$,  and:
\bea
\label{cw<1}
c_i^< &=&  \int_{\lambda_{min}}^{ \tilde{\omega}^{1/2}}\frac{ d\lambda}{\lambda}  \int_0^{2\pi} d\phi~ g_i(\phi) e^{- 8 i  \lambda \hat{\delta}_s \cos(\phi - \psi)} \nonumber \\
&\times&  \left[ e^{4 i \lambda^2 \cos(2 \phi)} - 1 \right]\, ,
\eea
with  $g_+ = \cos(2 \phi)$, $g_{\times} = \sin(2 \phi)$ and $\lambda_{min} = \tilde{\omega}^{1/2} w_{min}$.

$c_i^<$ depends non trivially on $ \hat{\delta}_s$ and $\psi$ and depends on $ \tilde{\omega}$ only through the upper limit of integration over $\lambda$. Then, by differentiating $c_i^<$ with respect to $ \tilde{\omega}$  we find that this latter dependence goes away at large $ \tilde{\omega}$ unless $\delta_s \le \tilde{\omega}^{-1}$, a region suppressed by phase space. In other words, the integral in $\lambda$  converges to a finite function of   $ \hat{\delta}_s$ and $\psi$ with subleading corrections in $\tilde{\omega}^{-1}$.  We also find that $c_i^<(\psi,  \hat{\delta}_s)$ is smooth and regular in $\psi$, goes to a constant at small $ \hat{\delta}_s$, and decreases like $ \hat{\delta}_s^{-3/2}$ at large $ \hat{\delta}_s$. As a consequence, when inserted into (\ref{dEGW}), the integral over $ \hat{\delta}_s$ and $\psi$  gives a finite result dominated by the region $ \hat{\delta}_s \sim 1$. Thus the resulting "scale-invariant" ($\omega^{-1} d \omega$) spectrum can be understood as coming from a solid angle shrinking like $\omega^{-1}$ for $\omega > 1/E$.

Before discussing the implications of this result we shall  argue that the region $w >1$ cannot change this conclusion. For $w >1$ we can use the large-$w$ approximation for $\Phi$. Defining now $\lambda =  \tilde{\omega} w$ we find:
\bea
\label{cw>1}
c_i^> &=& \int_{ \tilde{\omega}}^{ \infty}\frac{ d\lambda}{\lambda}  \int_0^{2\pi} d\phi~ g_i(\phi) e^{- 8 i  \lambda \delta_s \cos(\phi - \psi)} \nonumber \\
&\times& \left[ e^{-8 i \lambda \cos( \phi)} - 1 \right]\, .
\eea
A reasoning very close to the one used above shows that this integral has a very weak $\tilde{\omega}$ dependence at large $\tilde{\omega}$ so that the entire integral is strongly suppressed.

In conclusion we have found that the GW spectrum at $\omega \gg  1/E$ is of the $d \log \omega$ type with a coefficient that can be computed (numerically?) from  (\ref{dEGW}) and (\ref{cw<1}) with their respective upper limits of integration removed. Such a spectrum would lead to a divergent result if integrated over $\omega$ up to $\infty$. However,
 the presence of a lower cutoff in  $\lambda $ in (\ref{cw<1}) cuts off the spectrum at $\omega \sim E^{-1} \theta^{-2}$ (i.e. when the lower cut off in $\lambda$ becomes larger than 1) giving for the total GW yield from the high-frequency part of the spectrum:
\be
\label{total}
\frac{E^{GW}}{E}(\omega > E^{-1})  \sim  \kappa ~ \theta^2 \log(\theta^{-2})
\ee 

At this point we can go back to our simplified argument of Section III and justify the approximation (\ref{delta}) in the relevant kinematical region. Let's assume the validity of (\ref{EGW}) up to some maximal value $\bar{\rho}$ of $\rho_s$. Performing the $\rho_s$ integration and going over to the dimensionless  variables $w, \tilde{\omega}$, we easily get:
\bea
&& \frac{1}{E} \frac{d E^{GW}}{d \tilde{\omega}} = \frac{\bar{\rho}^2}{ \pi^3} \int \frac{d^2 w_1}{w_1^2} \frac{d^2 w_2}{w_2^{*2} }  \left( e^{-iE \omega \Phi(w_1)}-1\ \right)  \times\nonumber \\
&& F_{01}\left(2, - \frac{b^2}{4E^2} \bar{\rho}^2 \tilde{\omega}^2 |w_1-w_2|^2\right)\! \left( e^{+ iE \omega \Phi(w_2)}-1\ \right) 
\eea
For $\tilde{\omega} >1$ we can safely assume that $b^2 E^{-2} \bar{\rho}^2 \tilde{\omega}^2 \rightarrow \infty$ as $\theta \sim E/b \rightarrow 0$. As a result, the confluent hypergeometric function $F_{01}$ approaches $4 \pi E^2 b^{-2}\bar{\rho}^{-2} \tilde{\omega}^{-2} \delta^{(2)}(w_1-w_2)$ exactly reproducing (\ref{spectrum}).

\vspace{2mm}
\centerline{{\bf B: $ \theta <  \tilde{\omega} < 1$ i.e. $1/b < \omega < 1/E$}.}

In this case we have to distinguish, a priori, three integration ranges in $w$: $w <1$, $ 1<w < \tilde{\omega}^{-1}$ and $w > \tilde{\omega}^{-1}$. 
 In the first region we can  expand the square bracket of (\ref{cw<1}) and obtain (with $\lambda = \tilde{\omega} w$):
 \be
 c_i^< = \frac{i \pi}{16 \tilde{\omega}\delta_s^2}\int_0^{8 \tilde{\omega} \delta_s} d\lambda \lambda  \int_0^{2\pi} d\phi~ g_i(\phi)  \cos(2 \phi) e^{-  i  \lambda  \cos(\phi - \psi)} \, ,
 \ee
The integral over $\phi$ can be done explicitly in terms of Bessel functions and, as a result, $c_i^<  \sim \tilde{\omega}$ at small $ \tilde{\omega} \delta_s$ while it goes like $\tilde{\omega}^{-1/2} \delta_s^{-3/2}$ at large $ \tilde{\omega} \delta_s$. Inserting these behaviors in (\ref{EGW1}) leads to a flat spectrum in $\omega$.

Consider now the integration range $ 1<w < \tilde{\omega}^{-1}$. Here we have to use the large-$w$ approximation to $\Phi$ but we can still expand the exponential in  the square bracket of (\ref{cw>1}) and write (with $\lambda =\tilde{\omega}\delta_s w$):
\be
\label{cw>1'}
c_i^> = \frac{8 }{i \delta_s} \int_{ \tilde{\omega}\delta_s}^{ \delta_s} d\lambda  \int_0^{2\pi} d\phi~ g_i(\phi) \cos( \phi) e^{- 8 i  \lambda  \cos(\phi - \psi)}  \, .
\ee
Also this integral over $\phi$ leads to some combination of Bessel functions. Taking that into account we find that the region $\delta_s <1$ is strongly suppressed, while the region $\delta_s > 1$ gives its leading contribution from the sub-interval $1 < \delta_s < \tilde{\omega}^{-1}$ in which $\lambda$-integration from $\delta_s \tilde{\omega}$ to 1 gives $c_i^>  \sim \delta_s^{-1}$. Such a $c_i$ when inserted into (\ref{EGW1}) gives:
\be
dE^{GW}/d \omega \sim  \int_1^{ \tilde{\omega}^{-1}} \ \frac{d \delta_s}{ \delta_s} \sim  \log ( \tilde{\omega}^{-1})\, ,
\ee 
 while other integration regions cannot develop such a logarithmic enhancement. Note that at $ \tilde{\omega} = \theta$ the upper limit of integration over $\delta_s$ becomes of order $\theta^{-1}$ which is its maximal possible value (corresponding to $\rho \sim 1$).
 
 We are finally left with the range $w > \tilde{\omega}^{-1}$.  As before, in this range we may neglect the exponential inside the square brackets of (\ref{cw>1}). This leads again to a fast-converging integral dominated by the lower end of the integration range. The contribution is a constant spectrum with no logarithmic enhancement.
 We conclude that, at $1/b < \omega < 1/E$ the spectrum is flat up to a $\log (E \omega)^{-1}$ which arises from the angular region $\theta < \rho_s < (\omega b)^{-1}$.

 Thus, as we approach $\omega \sim 1/b$, the  $\log (E \omega)^{-1}$ becomes a  $\log \theta^{-1}$ enhancement  in  agreement with our previous discussion on the zero frequency limit.
 Actually the agreement is quantitative. In this low-frequency limit the integral over $\lambda$ in (\ref{cw>1'}) can be extended from 0 to $\infty$ and, going back to to complex notation for $c$, we find:
 \bea
 \label{ZFLc}
 && c \rightarrow  \frac{2 i \pi}{\delta_s} e^{2 i \psi} \int_0^{\infty} da  \left( \cos \psi  J_1(a) - 2 e^{ i \psi} \frac{J_2(a)}{a}  \right)  \nonumber \\
 &=& \frac{2  \pi}{\delta_s} e^{2 i \psi} \sin \psi \, , 
 \eea
 where we have used some known integrals involving Bessel functions.
 Finally, inserting (\ref{ZFLc}) into (\ref{EGW1}) we get:
 \be
 \frac{d E^{GW}}{d \omega} =  \frac{E^2 \theta^2}{ \pi^4} ~\int _1^{\bar{\delta}_s} \frac{d^2  \delta_s}{\delta_s^2} 4 \pi^2 \sin^2 \psi \rightarrow \frac{4}{\pi} E^2 \theta^2 \log \theta^{-1}\, ,
 \ee
 in full agreement with (\ref{lowom}) and (\ref{ZFL}). 
 If instead we integrate the high-frequency spectrum (\ref{intermediate}) up to the postulated cutoff $\omega_{max} =  \theta^{-2} E^{-1}$ we obtain, to logarithmic accuracy,
 \be
 \frac{E^{GW}}{E}  = \frac{1}{ 2 \pi} \theta ^2 \log \theta^{-2}\; .
 \label{efficiency}
 \ee

\section{Conclusions}

We have proposed a simple approach to the problem of GW radiation from the collision of massless particles obtaining very sensible results. In particular, we were able to reproduce  the correct zero-frequency-limit  and obtained an almost flat ($\sim \log (E \omega)$) spectrum up to $\omega \sim E^{-1}$.
The simplified treatment of Section III, as well as the more detailed analysis of Section IV, support the conclusion that there is a break in the frequency and angular distribution at this characteristic  frequency. The former becomes scale-invariant (i.e. $d E^{GW}/d \omega \sim \omega^{-1}$) with the GW emitted at smaller and smaller angles (with respect to the deflected trajectories) with increasing frequency ($\rho \sim \theta (\omega E)^{-1/2}$).

 Such a spectrum implies a logarithmic sensitivity of the total GW yield to the high-frequency cutoff. Looking at the characteristic length scales involved suggests setting the UV cutoff at $\omega = \omega_{cr} \sim  \theta^{-2}  E^{-1 }$.
 The necessity of such a cutoff is also supported by a conjectured upper bound on the power emitted in GW. It is also in agreement with  the naive  conclusions of \cite{ACV, VW} on the analog quantum problem.  However, while in  \cite{ACV}, \cite{VW}  the spectrum was assumed to be constant up to the cutoff,  with rather paradoxical consequences, 
 here the scale-invariant spectrum leads  to a much more reasonable efficiency  for GW production of order  $\theta ^2 \log \theta^{-2}$ (with a precise numerical coefficient)  in qualitative agreement (i.e. modulo the logarithmic enhancement) with previous guesses \cite{guess} based on a rough estimate of the interaction time. 
 
 Meanwhile an argument providing the same large-$\omega$ suppression in the quantum problem has been given \cite{CCV}. It appeals to the fact that graviton emission from different rungs in the ladder responsible for the leading eikonal is only coherent over a fraction $\Delta \theta/\theta \sim (\omega b \theta)^{-1} \sim  (\omega E)^{-1}$ of the whole ladder. Even more recently, it was shown \cite{CCCV} that combining the previous effect with the modified rescattering phase of the graviton production amplitude reproduces, modulo quantum corrections, {\it exactly} our classical result (\ref{differential}).

We ought to conclude with two less positive remarks. On the phenomenological side it is difficult to imagine processes giving rise to a sufficient amount of GW to be relevant for direct, or even indirect, detection. An exception could be the GW yield during a very high-curvature phase in the early Universe. 
On the theoretical side, our method has limitations since it can only deal effectively with some  regions in GW phase space (frequency and angles). Although we believe that those regions are the most relevant ones for GW emission, we cannot exclude  that other kinematical configurations will provide comparable, or even higher, GW yields. 
In order to completely settle this issue a more general framework, like the one of \cite{Grisha}, appears to be necessary but goes beyond the aims of this first study of the problem.
\section*{Acknowledgements}
We have benefitted from many discussions with colleagues at NYU's CCPP  and, in particular, with Sergei Dubovsky, Gia Dvali, Victor Gorbenko and Hovhannes Grygorian.
We would also like to acknowledge interesting correspondence with Vitor Cardoso, Rafael Porto, Frans Pretorius and Theodore Tomaras. One of us (GV) would like to thank Marcello Ciafaloni, Dimitri Colferai and Francesco Coraldeschi for very useful discussions and for pointing out a trivial  numerical error in the original version of this paper. He also wishes to thank Thibault Damour and Robert Wald for discussions and Grisha Vilkovisky for constructive criticism and  informative correspondence on the radiation-moment approach to gravitational radiation.

\end{document}